\documentclass[reprint,amsmath,amssymb, aps,]{revtex4-2}
\usepackage{CJK}

\usepackage[english]{babel}
\usepackage[utf8]{inputenc}
\usepackage{graphicx}
\usepackage{amssymb}
\usepackage{amsthm}
\usepackage{color}
\usepackage{bm}
\usepackage{braket}
\usepackage{setspace}
\usepackage{subfigure}
\usepackage{float}
\usepackage{comment}

\usepackage{graphicx}
\usepackage{dcolumn}
\usepackage{bm}
  
\begin{document}

\preprint{APS/123-QED}

\title{Broadband Infrared Study of Pressure-Tunable Fano Resonance and Metallization Transition in 
2H-MoTe$_2$.}

\author{E. Stellino} 
\email{elena.stellino@studenti.unipg.it}
\affiliation{Department of Physics and Geology, University of Perugia, via Alessandro Pascoli, Perugia, Italy}

\author{F. Capitani}
\email{francesco.capitani@synchrotron-soleil.fr}
\affiliation{Synchrotron SOLEIL, L’Orme des Merisiers, Saint-Aubin, Gif-sur-Yvette, France}

\author{F. Ripanti}
\affiliation{Department of Physics, Sapienza University of Rome, P.le A. Moro 5, Rome, Italy}

\author{M. Verseils}
\affiliation{Synchrotron SOLEIL, L’Orme des Merisiers, Saint-Aubin, Gif-sur-Yvette, France}

\author{C. Petrillo}
\affiliation{Department of Physics and Geology, University of Perugia, via Alessandro Pascoli, Perugia, Italy}

\author{P. Dore}
\affiliation{Department of Physics, Sapienza University of Rome, P.le A. Moro 5, Rome, Italy}

\author{P. Postorino}
\affiliation{Department of Physics, Sapienza University of Rome, P.le A. Moro 5, Rome, Italy}

\date{\today}

\begin{abstract}
High pressure is a proven effective tool for modulating inter-layer interactions in semiconducting transition metal dichalcogenides, which leads to significant band structure changes. Here we present an extended infrared study of the pressure-induced semiconductor-to-metal transition in 2H-MoTe$_2$, which reveals that the metallization process at 13$\div$15 GPa is not associated with the indirect band gap closure, occuring at 24 GPa. A coherent picture is drawn where n-type doping levels just below the conduction band minimum play a crucial role in the early metallization transition. Doping levels are also responsible for the asymmetric Fano line-shape of the E$_{1u}$ infrared-active mode, which has been here detected and analyzed for the ﬁrst time in a Transition Metal Dichalcogenide compound. The pressure evolution of the phonon profile under pressure shows a symmetrization in the 13$\div$15 GPa pressure range, which occurs simultaneously with the metallization and confirms the scenario proposed for the high pressure behaviour of 2H-MoTe$_2$.
\end{abstract}

\maketitle

\section{Introduction}

Transition Metal Dichalcogenides (TMDs) are layered crystals with stoichiometry MX$_2$, where M is a transition metal atom and X is a chalcogen element \cite{Manzeli,Kolobov}. Since the intra-layer covalent bonds are far stronger than the inter-layer van der Waals forces, few-layer down to mono- layer samples can be exfoliated from the bulk crystal, obtaining almost two-dimensional systems \cite{Li,Yuan}.

In molybdenum-based TMD semiconductors, the lattice anisotropy in the out-of-plane direction is responsible for the connection between the electronic properties and the number of layers. A progressive increase in the indirect band-gap is observed as the number of layers is reduced, and an indirect-to-direct band-gap crossover arises when bi-layer samples are scaled down to mono-layers  \cite{Yun, Zhang, Mak2}. 
In particular, 2H-MoTe$_2$ (hereinafter referred to as MoTe$_2$) exhibits an indirect band-gap E$_g^I\sim$ 0.9 eV and a direct gap  E$_g^D \sim$1 eV in the mono-layer \cite{Wang}. Compared to the most widely studied 2H-MoS$_2$ and 2H-MoSe$_2$, whose band-gaps lie in the visible region \cite{Wang}, the smaller MoTe$_2$ energy gap makes it suitable as optoelectronic device also in the near-infrared range \cite{Pradhan, Huang, Yu}. 

Tuning the electronic properties through the modulation of the inter-layer interaction can further boost the technological applications for MoTe$_2$ crystals. This is typically achieved by changing the sample thickness and studying the band structure dependence on the number of layers. A complementary and quite clean approach that does not require changing the sample is given by the high pressure application, which allows for the analysis of the crystal response when the inter-layer bonds strengthen.

Previous studies based on DFT calculations show that the MoTe$_2$ lattice symmetry is maintained up to 50 GPa \cite{Kohulak}; this prediction is in agreement with high-pressure Raman findings \cite{zhao, Bera, Yang, Stellino2020} up to $\sim$ 30 GPa, with the number of peaks unchanged and their position continuously shifting towards higher frequencies on increasing pressure.
As to the electronic properties, theoretical and experimental works indicated the occurrence of a pressure-induced metallization although the transition pressure value is still controversial. In particular, DFT calculations by Riflikova et al. placed the transition in the range 13-19 GPa \cite{riflikova}.
The Raman measurements of Bera et al. suggested an electronic transition around 6 GPa \cite{Bera}. Finally, recent resistivity measurements provided evidence for the metallization above $\sim$10 GPa \cite{zhao} and, specifically, at $\sim$15 GPa \cite{Yang}. In these two papers, DFT calculations were also performed, showing a band gap closure at $\sim$10 GPa \cite{zhao} and $\sim$13 GPa \cite{Yang}. Although all the above mentioned works implicitly associated the onset of the metallization process with the indirect band-gap closure, none of them provided a direct experimental determination of the E$_g^I$ pressure dependence.
\\
Within this still debatable picture, a direct investigation of the electronic properties of MoTe$_2$ is needed to accurately characterize the nature of the observed transition and identify the metallization pressure. Infrared spectroscopy is the ideal technique to investigate the evolution of the TMD electronic transitions under pressure and it represents the most suitable alternative to photoemission studies that are impossible at such high pressures.

We herein report the first synchrotron-based infrared absorption measurements in MoTe$_2$ as a function of pressure.
The high-pressure transmission experiments were performed at synchrotron Soleil at room temperature (RT) exploring two frequency ranges: the near-infrared (NIR) at the SMIS beamline, and the far-infrared (FIR) at the AILES beamline.
The NIR experiment allowed us to measure the indirect gap over the 0$\div$27 GPa range, revealing its closure and thus the occurrence of the transition to a metallic band structure at 24 GPa. In the FIR range, we found an abrupt increase of the free-carriers related spectral weight (Drude contribution) at  $\sim$13 GPa, which can be identified as the signature of the metallization process previously observed by resistivity measurements \cite{zhao}.

Investigating the pressure-induced changes of the electronic properties in MoTe$_2$ over two distinct frequency domains enabled us to measure both the indirect energy gap and the variation of the charge carrier density as a function of pressure. That allowed us to frame the metallic transition in a coherent scenario where the contribution from the doping levels present in the band structure is properly taken into account. Moreover, the asymmetric E$_{1u}$ profile observed at low pressures, which abruptly symmetrizes as the carrier density increases at P=13$\div$15 GPa, not only proves itself as a new example of Fano resonance in the TMD class of systems, but it also contributes supporting the complex picture we believe is behind the pressure-driven metallization in MoTe$_2$.

\section{Experimental}

A MoTe$_2$ crystal, provided by HQ Graphene, was exfoliated to obtain fresh-cut samples. RT, high-pressure infrared transmission measurements were carried out at the synchrotron SOLEIL. 0$\div$27 GPa pressure range was achieved by using a Diamond Anvil Cell (DAC). Diamonds with culets of 400 $\mu$m  were separated by a pre-indented stainless steel 50 $\mu$m thick gasket, in which a hole of 150 $\mu$m diameter was drilled. The exfoliated sample was positioned in the hole together with the pressure transmitting medium and a ruby chip, to measure the pressure through the ruby fluorescence technique. 
The measured samples typically had a 50x50 $\mu$m$^2$ surface and micrometric thickness (bulk crystals). 

NIR measurements were performed over the 4000$\div$9000 $cm^{-1}$ range at the SMIS beamline, using a Thermo Fisher iS50 interferometer equipped with a quartz beamsplitter. Synchrotron edge radiation was employed as a light source.
Custom-made 15x Cassegrain objectives with a large working distance allowed focusing and then collecting the transmitted radiation, which was finally detected by a liquid-nitrogen cooled mercury-cadmium-telluride detector \cite{Livache}. NaCl powder was used as the hydrostatic medium \cite{Celeste}.

FIR measurements were carried out by using the setup available at the AILES beamline \cite{roy, voute}. The Bruker IFS 125 HR interferometer, coupled with the synchrotron source, was equipped with a multi-layer 6 $\mu$m beamsplitter and a 4K-bolometer. Polyethylene powder was used as a pressure transmitting medium \cite{Zhukas} allowing for reliable measurements over the 100$\div$600 cm$^{-1}$ spectral range. 

In both FIR and NIR experiments, the absorbance spectrum $A(\omega)=-ln[I(\omega)/I_0(\omega)]$ was obtained at each pressure, with $I(\omega)$ the intensity transmitted with the MoTe$_2$ sample loaded in the cell and $I_0(\omega)$ the background intensity with the DAC filled by the hydrostatic medium only.

Finally, ambient condition measurements in the FIR range were performed by a Nicolet NicPlan microscope (32x magnification, 100x100 $\mu m^2$ area) coupled with synchrotron radiation as light source and liquid He bolometer as a detector. In this case, larger samples with $\sim$ 200x200 $\mu m^2$ area were measured. 

\section{NIR measurements}

Room temperature NIR transmission spectra were collected over the 0$\div$17 GPa and 0$\div$27 GPa pressure ranges on two  MoTe$_2$ samples with different thickness: $\sim$10 $\mu m$ (\textit{first} sample) and $\sim$3 $\mu m$ (\textit{second} sample). Raw spectra show clear interference fringes at low pressure characterized by an amplitude that decreases as pressure and absorption increase. This behaviour, as well as the fringes frequency spacing, is consistent with the occurrence of multiple reflections between the faces of the sample slab. Applying a simple Fourier spectral smoothing procedure, the interference fringes are consistently reduced, which emphasizes the sample absorption profile.

\begin{figure*}[ht]
\centering
  \includegraphics[width=1\textwidth]{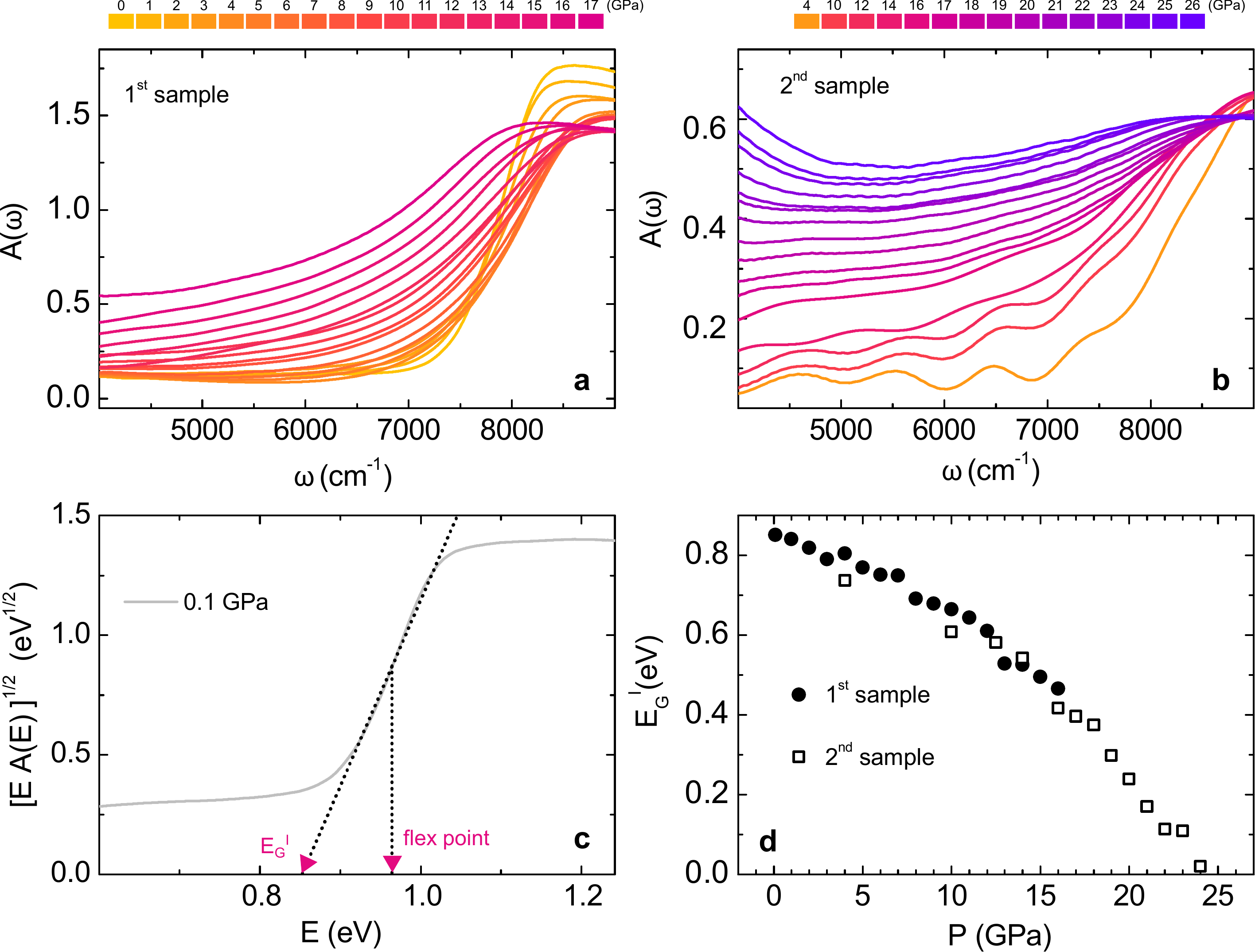}
  \caption{(a) Absorbance spectra collected in the first run over the [0,1 $\div$ 17] GPa range. (b) Absorbance spectra collected in the second run  over the [4 $\div$ 27] GPa range. (c) Tauc-plot at $P=0.1$ GPa and linear extrapolation of the energy gap. (d) Energy gap as a function of the pressure for the first (dots) and the second (open squares) run.}
  \label{fig1}
\end{figure*}

The absorbance spectra, $A(\omega)$, obtained after the smoothing procedure are shown in Fig. \ref{fig1}(a),(b) for the \textit{first} and the \textit{second} sample respectively. The absorbance exhibits a sigmoidal shape, which is well defined at P=0 GPa and flattens as the pressure increases. A tail develops below 5000 cm$^{-1}$ above $\sim$15 GPa and it increases with pressure, suggesting the formation of a broad band below the lower limit of the probed IR range. At the same time, the inflexion point of the sigmoidal curve shifts rather continuously towards the low-frequency side of the spectrum.

These features are related to the pressure evolution of MoTe$_2$ electronic properties. Indeed, the steep rise of the absorbance at high frequencies (above about 7000 cm$^{-1}$) can be ascribed to electronic transitions from valence to conduction band \cite{brotons2018, Lezama, zhao2}, while the aforementioned tail at low frequencies can be attributed to a Drude-like contribution originating from the progressive increase of the charge carrier density in the conduction band \cite{Postorino2003, Calvani1998, Stellino2021}. As mentioned in the introduction, in MoTe$_2$ both the direct and indirect band gaps lie within the NIR energy range here explored (i.e. about 0.4$\div$1.2 eV)  and the energy difference between the two gaps is $\sim$ 0.1 eV ($\sim$ 800 cm$^{-1}$), significantly smaller than in other TMD semiconductors. In principle, both the transitions could be visible in the measured spectra.
However, as already observed by Lezama et al. \cite{Lezama}, when the crystal thickness is larger than 1 $\mu$m, the high absorption due to the indirect transition strongly reduces the transmitted intensity above E$_g^I$, which prevents the observation of the direct optical transition at the higher energy E$_g^D$. We also note that, in the present case, the transmitted intensity is further reduced by the reflectivity of the diamond anvils faces, which makes it even more difficult to reveal the signature of the direct transition. Since the thickness of our samples is larger than 1$\mu$m (in particular, the \textit{first} sample thickness is d$\sim$ 10 $\mu m \gg 1 \mu m$) we can safely ascribe the steep rise observed in $A(\omega)$ above 7000 cm$^{-1}$ to the indirect-gap transition.

Under the assumption of an indirect gap transition, the absorption coefficient, $\alpha(\omega)$, in the neighborhood of E$_g^I$ can be written as:
$\sqrt{\alpha(\omega)\cdot\hbar\omega}\propto \hbar\omega-E_g^I \pm E_{ph}$,
where $E_{ph}$ is the energy of the phonon required by the momentum conservation. Under the reliable assumption that the measured $A(\omega)$  is simply proportional to $\alpha(\omega)$ and neglecting the phonon energy $E_{ph}$, E$_g^I$ can be determined by using the so-called Tauc plot. This amounts to exploiting a linear regression for the quantity $\sqrt{\alpha(\omega)\cdot\hbar\omega}$ \cite{Tauc,Viezbicke,zhao2}, as shown in Fig. \ref{fig1}(c). The Tauc plot procedure was applied to all the experimental absorption spectra. We note that the energy gap at 0 GPa is 0.87 eV, in perfect agreement with the 0.88 $\pm$ 0.05 eV value recently observed in infrared measurements at ambient conditions \cite{Lezama}. The resulting E$^I_g$ values are shown in Fig. \ref{fig1}(d) as a function of pressure for both the considered samples. The two sets of measurements are in good agreement over the common pressure region and show a progressive reduction of the band-gap that finally closes at P $\sim$ 24 GPa. The tail of the Drude-like band appears at  $\sim$15 GPa, although the gap closure was estimated around 24 GPa, which suggests that a significant increase of the free charge carriers in the conduction band occurs at a pressure where the indirect band-gap is still wide open.  

\section{FIR measurements}
\begin{figure*}[ht]
\centering
  \includegraphics[width=0.9\textwidth]{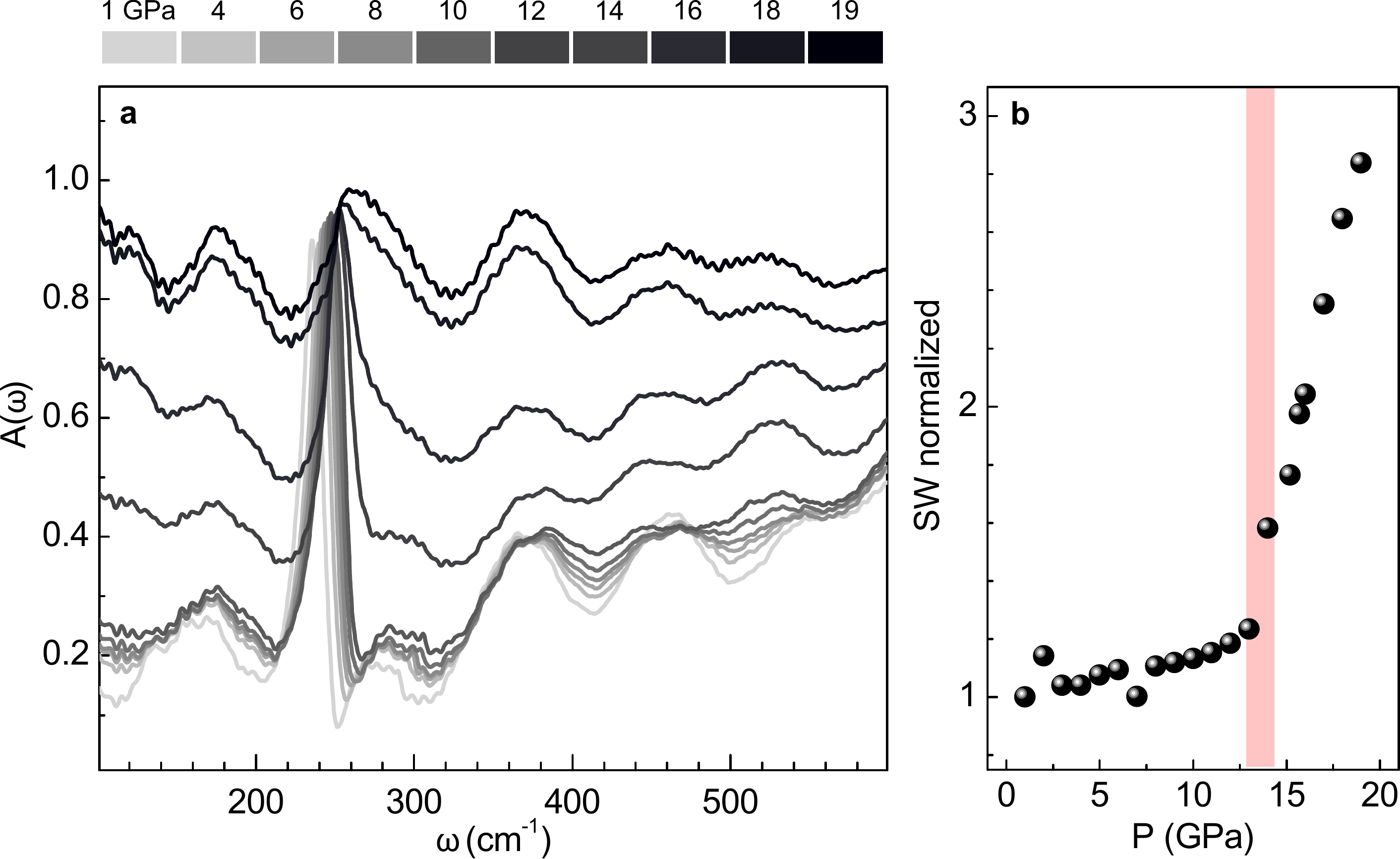}
  \caption{(a) Absorbance spectra at selected pressure values over the range [1 $\div$ 19] GPa. (b) Normalized spectral weight $SW(P,P_0)$ (see text); the red vertical stripe highlights the pressure range where the change of slope occurs.}
  \label{fig2}
\end{figure*}

To better characterize the pressure response of the Drude contribution in the sample absorbance, we collected $A(\omega)$ spectra in the FIR region over the [1$\div$19] GPa range, see Fig. \ref{fig2}(a). At every pressure, both $I_0(\omega)$ and $I(\omega)$ exhibit intense oscillations, clearly visible also in the corresponding $A(\omega)$ curve. In this case, measurements were collected in the inter-band energy region where the MoTe$_2$ is largely transparent and the interference fringes are compatible with multiple reflections between the diamond inner surfaces through the polyethylene medium. Although a simple Fourier spectral smoothing procedure was not completely effective to remove the interference fringes, a sharp peak at about 235 cm$^{-1}$ at low pressure was quite apparent and it was assigned to the E$_{1u}$ phonon according to literature \cite{wieting, caramazza}. The phonon peak also shows a clear broadening and a frequency hardening on increasing pressure, an effect that will be discussed in detail in the next section. Focusing on the overall trend of the spectra in Fig. \ref{fig2}(a), it can be noticed that they remain almost overlapped at low pressures (approximately over 0$\div$13 GPa), while their intensities rapidly rise on further increasing pressure. A simple application of the Drude model proves that in the low-frequency limit, i.e. $\omega/\Gamma \ll 1$ with $\Gamma=c/\tau$ and $\tau$ the Drude relaxation time, the experimental absorbance is simply proportional to the square root of both the DC conductivity $\sigma_0$ and the frequency $\omega$, that is $A(\omega) \propto \sqrt{\sigma_0 \omega}$. Since $\sigma_0$ is proportional to the free carrier density n, an increase in the FIR absorbance is directly related to an increase of n \cite{Stellino2021}. Therefore, the significant growth in the overall absorption at high pressure can be related to the onset of an early metallization process in the MoTe$_2$ sample. To carry out a quantitative analysis, we define the absorption spectral weight at each pressure:
$\displaystyle{sw(P)=\int_{\omega_m}^{\omega_M}{A(\omega)d\omega}}$.
We then conveniently introduce the normalized quantity $ SW(P,P_0)= sw(P)/sw(P_0)$, which describes the variation of the spectral weight from the minimum pressure value P$_0$ =1 GPa. In Fig. \ref{fig2}(b) we show the $SW(P,P_0)$ values obtained with $\omega_m$ =100 cm$^{-1}$ and $\omega_M$ =600 cm$^{-1}$. The integration does not include the 200$\div$270 cm$^{-1}$ range where the phonon contribution is dominant. Our results indicate the onset of a phase with metallic characteristics between 13$\div$15 GPa, in good agreement with previous resistivity measurements \cite{zhao, Yang}. This finding is also consistent with the onset of the low-frequency contribution (Drude peak), as discussed in the previous section. 
We want to finally notice that the spectral weight analysis is a robust procedure as the same trend of the spectral weight function with pressure (Fig. \ref{fig2}(b)) is obtained either non excluding the phonon frequency region or varying the integration limits $\omega_m$ and $\omega_M$ (data not shown). 

\section{The Fano resonance}
\begin{figure*}[ht]
\centering
  \includegraphics[width=1\textwidth]{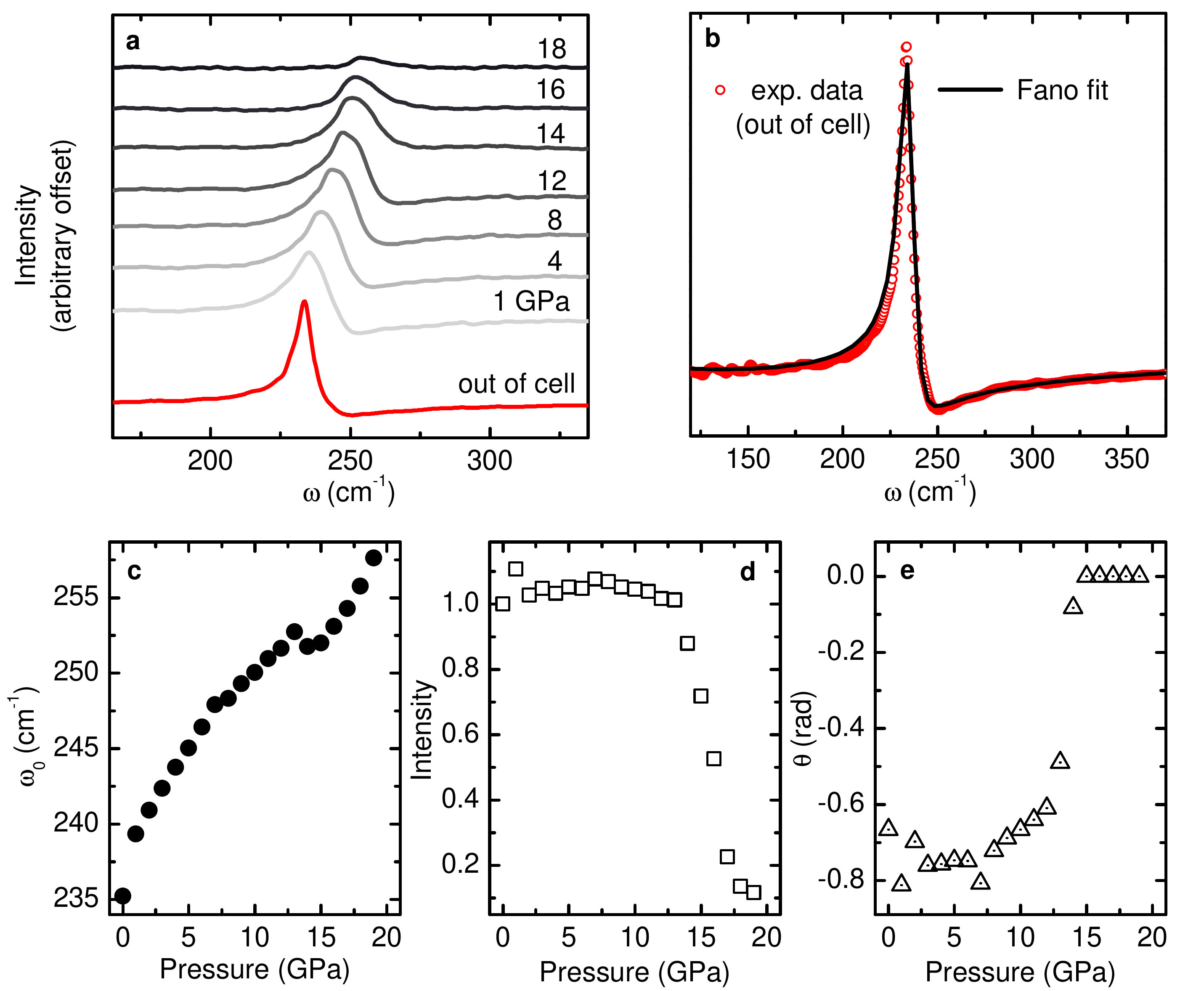}
  \caption{(a) Absorbance spectra at selected pressures after the subtraction of a parabolic background and of the sinusoidal fringes. Spectra are vertically shifted for clarity. The red spectrum has been collected at ambient conditions, out of the DAC, as a reference. (b) Experimental data at ambient pressure (red dots) fitted by a Fano function (black line). (c) E$_{1u}$ peak position as a function of pressure. (d) Peak intensity as a function of pressure, normalized to the 1 GPa value. (e) $\theta$ parameter of the Fano fit as a function of pressure.}
  \label{fig3}
\end{figure*}

As mentioned, FIR absorption spectra exhibit an intense phonon peak at 235 cm$^{-1}$ attributed to the E$_{1u}$ vibrational mode. The inspection of the raw spectroscopic data shown in Fig. \ref{fig2}(a) reveals that the E$_{1u}$ peak presents an unusual asymmetric shape in the low pressure regime, and it is notably affected by pressure. To rule out the effect of the diamond-diamond interference fringes on the line shape asymmetry, a further absorption measurement was carried out on a simple MoTe$_2$ slab kept out of the DAC (that is at P=0 GPa and without interference effects). The line shape of the out-of-cell phonon peak, shown in figure \ref{fig3}(a), displays a remarkable Fano-like profile \cite{Homes} in close similarity with what was observed in the inside-cell measurements.

The measured phonon peak line-shape was modelled by using a Fano-profile \cite{Fano} with a second-order polynomial to fit the background. The Fano profile is given by:  

\begin{equation} \label{eq}
F(\omega)=F\cdot\frac{\omega^2\cos\theta+\omega(\omega^2-\omega_0^2)\sin\theta}{\omega^2\gamma^2+(\omega^2-\omega_0^2)} 
\end{equation}

where $\omega_0$ is the peak centre, $\gamma$ is the peak width and $\theta$ is the phase associated with the line asymmetry. Notice that $\theta$ is directly related to the well-known Fano parameter $q$  by $\tan(\theta/2)=1/q$ \cite{Fano, Homes} with, however, the numerical advantage of avoiding the $q$ divergence. The symmetric Lorentzian profile, corresponding to the no resonance condition, is indeed recovered when $\theta=0$ (i.e. $q \rightarrow \infty$). In figure \ref{fig3}(b) we can appreciate the very good agreement between the best fit curve and the experimental data a P=0. Best fit parameters are plotted in panels c-e of Fig.\ref{fig3} at P=0.

This result validates a fitting procedure of the pressure-dependent spectra, which is based on the combination of three functions: a second-order polynomial background, a Fano-profile for the E$_{1u}$ phonon, and a sinusoidal curve for the interference fringes. In fig. \ref{fig3}(a) the spectra collected at selected pressure values after electronic background and fringes subtraction are shown. 

The best fit peak parameters of the Fano profiles are shown in figures \ref{fig3}(c),(d),(e) as a function of pressure. The peak central frequency $\omega_0$ (figure \ref{fig3}(c)) shifts toward higher values on increasing pressure, as expected. We remind that up to 50 GPa no evidence for a structural transition has ever been found, neither theoretically \cite{Kohulak} nor experimentally \cite{zhao, Bera, Yang}. Therefore, the weak anomaly observed at around 13 GPa in the pressure dependence of the peak frequency $\omega_0$ could be ascribed to some electronic effect. Correspondingly, the E$_{1u}$ intensity (figure \ref{fig3}(d)) remains almost constant up to 13 GPa and then rapidly decreases. This is interpreted as a consequence of the optical response of free electrons that screen the phonon contribution in the FIR absorption. As to the peak profile, figure \ref{fig3}(e) clearly shows that the $\theta$ parameter keeps the almost constant value $\sim$0.7 rad up to 13 GPa, and then abruptly goes to zero as the peak symmetrizes and the Fano coupling disappears. 
The overall line shape analysis thus consistently points out at the presence of a pressure threshold of electronic nature above 13 GPa. 

The Fano theory \cite{Fano} explains the asymmetric profile of a given discrete excitation as a consequence of the coupling between the excitation itself and a continuum of excited states with the same energy scale. In the present case, the discrete state is undoubtedly the E$_{1u}$ phonon, with energy E$_{ph}$=0.03 eV whereas the continuum needs to be identified by some suitable electronic transitions with the appropriate energies. The MoTe$_2$ band gap is E$_g$ $\sim$ 0.9 eV, that is more than one order of magnitude larger than E$_{ph}$, and the phonon coupling to a transition from valence to conduction band cannot take place. However, since the measured samples are n-type semiconductors, as declared by the manufacturer HQ-Graphene and like most of the TMDs grown by Chemical Vapor Transport \cite{zhao, Pisoni}, charges in excess giving rise to extra electronic levels, just below the conduction band minimum, could be responsible for electronic excitations with energies comparable with E$_{ph}$, as also reported in \cite{tarasov2015}. It is worth noticing that a negative asymmetry parameter $\theta$, as obtained from the experimental data fit, is perfectly consistent with the presence of n-type doping that is known to cause an asymmetric broadening of the peak on the lower energy side, as previously observed for few-layer graphene \cite{zhou2020, Cappelluti, Cappelluti2}. Conversely, p-type doping would have produced hole states in the proximity of the valence band maximum and an asymmetric broadening of the peak on the higher energy side. 
\\
In the present scenario, the pressure evolution of the phonon lineshape can be consistently explained: indeed, as the pressure increases, the energy separation between the conduction band minimum and the doping levels progressively reduces, until vanishing above 13-15 GPa. 
This process not only is responsible for the increase of the carrier density observed in the FIR in the high-pressure regime, but it also causes the observed phonon peak symmetrization. Indeed, when the energy gap between the doping level and the conduction band goes to zero, no more direct electronic transitions from inter-band states are available for the phonon to couple with, preventing the existence of the Fano resonance in the E$_{1u}$ profile.

\section{Discussion and conclusions}
The results of the different series of IR measurements at low (FIR) and high (NIR) frequency enabled a deeper understanding of the metallization process in bulk MoTe$_2$, unravelling the mechanisms that drive the transition and the role played by the doping levels. 
The interpretation model we propose largely reconciles the apparently controversial results that appeared in the literature on the metallization transition in MoTe$_2$, where the metallization pressure was reported to vary from 6 GPa \cite{Bera} to about 20 GPa \cite{riflikova}. Our experiment reveals that the spectroscopic signatures of the metallization are related to different microscopic mechanisms, which enable to frame the electronic transitions in a more complex scenario. 

The NIR measurements enabled us to identify the indirect nature of the transition and to obtain the band-gap energy as a function of the pressure. The pressure dependence of the energy gap revealed the closure of the band-gap at pressures above about 24 GPa, a value rather close to the upper limit of the pressure range indicated by the most detailed DFT calculations by M. Riﬂikova et al \cite{riflikova}.
DFT calculations focused on the study of the pressure behaviour of the electronic bands, identifying the occurrence of the metallization when the band overlap occurs. It is worth noticing that the effect of doping is not accounted for in DFT calculations and it is not detectable in the NIR experiment. 

In the FIR region, the spectral weight analysis of the measured absorbance showed an abrupt increase in the free carrier density in the conduction band at 13$\div$15 GPa. This finding is in very good agreement with the results of the resistivity measurements reported in the literature \cite{zhao, Yang}, which ﬁx the onset of the metallization in the 10$\div$15 GPa pressure range.
FIR measurements, as well as resistivity measurements, provide an estimate of the increase in the free carrier density in the conduction band as the pressure is increased. The presence of thermally excited doping-electrons in the conduction band significantly contribute to both the absorption spectral weight and the resistivity. The n-type doping thus induces a metallic behaviour before the effective overlap between conduction and valence band.

The relevance of the doping levels in the low energy electrodynamics is also witnessed by the Fano line-shape, here observed for the ﬁrst time in the E$_{1u}$ peak of MoTe$_2$. The occurrence of a Fano resonance indirectly provides an estimate for the energy gap between the conduction band minimum and the doping level in the band structure, which should be of the same order of magnitude as the E$_{1u}$ phonon, i.e. $\sim$ 0.03 eV. Moreover, the pressure evolution of the phonon line shape towards a symmetric Lorentzian profile at high pressure is consistent with the closure of the small energy gap exciton-conduction band. According to the Fano theory, the symmetrization of the E$_{1u}$ peak should occur when all the direct electronic transitions with energy comparable to E$_{ph}$ are suppressed. The peak symmetrization and the metallization onset are indeed observed at the same pressure.

In summary, our work provides the ﬁrst reliable measurement of the MoTe$_2$ indirect band-gap energy as a function of pressure, indicating a gap-closing pressure above 24 GPa. At the same time, we are able to provide an explanation for the early metallization observed in both FIR and electrical resistivity measurements by taking into account the doping effects. The pressure evolution of the Fano resonance in the E$_{1u}$ peak, here observed and interpreted for the first time, strongly supports the key role played by the doping levels in determining the sample electronic properties and sheds light on the pressure-induced modulation of the electron-phonon coupling in TMDs.  We finally underline that the present measurements clearly demonstrated the capability of infrared spectroscopy to provide deeper understanding of the electronic properties of TMDs either at ambient or under high applied pressure. 

\section*{Acknowledgements}
We acknowledge SOLEIL for provision of synchrotron radiation facilities (Proposal No. 20171166 on AILES beamline and Proposal No. 20191765 on SMIS beamline) and we would like to thank Professor Alessandro Nucara for assistance in interpreting the experimental results.

%\bibliographystyle{unsrt}
%\bibliography{biblio_new}

\end{document}